\documentclass[aps,pre,twocolumn,notitlepage,superscriptaddress]{revtex4-1}

\usepackage{amsmath}
\usepackage{amssymb}
\usepackage{hyperref}
\usepackage{graphicx}
\usepackage[arrow, matrix, curve]{xy}
\usepackage{bbm}
\usepackage{bm}
\usepackage{yhmath}
\usepackage{color}
\usepackage{physics}
\usepackage{bbold}

\usepackage{color}

\renewcommand\vec[1]{\boldsymbol{\mathrm{#1}}}

\newcommand\diff{\mathrm{d}}

\usepackage[usenames,dvipsnames]{xcolor}
\hypersetup{colorlinks=true, linkcolor=BrickRed, urlcolor=blue!50!black, citecolor=blue!50!black}

\usepackage[normalem]{ulem}
\makeatletter
\newcommand\hide@visible[1]{%
  \bgroup\fboxsep=.3ex\colorbox{Gray}{begin hide}%
  #1\colorbox{Gray}{end hide}\egroup%
}
\newcommand\hide@hidden[1]{%
  \bgroup\fboxsep=.3ex\colorbox{Gray}{hidden text}%
}
\newcommand\hide@invisible[1]{}
\newcommand\makevisible{\let\hide\hide@visible}
\newcommand\makehidden{\let\hide\hide@hidden}
\newcommand\makeinvisible{\let\hide\hide@invisible}
\makeatother
\makehidden

\begin{document}

\title{Active Brownian Particles in a Circular Disk with an Absorbing Boundary}

\author{Francesco Di Trapani}
\affiliation{Institut f\"ur Theoretische Physik, Universit\"at Innsbruck, Technikerstra{\ss}e 21A, A-6020, Innsbruck, Austria}
\author{Thomas Franosch}
\affiliation{Institut f\"ur Theoretische Physik, Universit\"at Innsbruck, Technikerstra{\ss}e 21A, A-6020, Innsbruck, Austria}
\author{Michele Caraglio}
\email{Michele.Caraglio@uibk.ac.at}
\affiliation{Institut f\"ur Theoretische Physik, Universit\"at Innsbruck, Technikerstra{\ss}e 21A, A-6020, Innsbruck, Austria}

\date{\today}

\begin{abstract}
We solve the time-dependent Fokker-Planck equation for a two-dimensional active Brownian particle exploring a circular region with an absorbing boundary.
Using the passive Brownian particle as basis states and dealing with the activity as a perturbation, we provide a matrix representation of the Fokker-Planck operator and we express the propagator in terms of the perturbed eigenvalues and eigenfunctions.
Alternatively, we show that the propagator can be expressed as a combination of the equilibrium eigenstates with weights depending only on time and on the initial conditions, and obeying exact iterative relations.
Our solution allows also obtaining the survival probability and the first-passage time distribution.
These latter quantities exhibit peculiarities induced by the non-equilibrium character of the dynamics, in particular, they display a strong dependence on the activity of the particle and, to a less extent, also on its rotational diffusivity.
\end{abstract}

\maketitle

\section*{Introduction}

In the past decades, active particles have received increasing attention because of their potential applications in various fields ranging from biology~\cite{Needleman2017,Lipowsky2005} to medicine~\cite{Henkes2020}, and robotics~\cite{Cheang2014,Erkoc2019}. 
Furthermore, because of their ability to convert energy into directed motion, active particles provide an ideal simplified model to investigate, from a statistical physics perspective, fundamental aspects of nonequilibrium systems~\cite{Cates2012,Romanczuk2012,Marchetti2013, Elgeti2015,Bechinger2016,Fodor2016}.

Nowadays, several aspects of active-particle dynamics have been successfully investigated by an impressive amount of experimental and theoretical research focusing on both single particles and their emerging collective behavior~\cite{Marchetti2013,Bechinger2016}.
However, exactly solvable models, allowing a deeper understanding on some basic theoretical aspects, remain rare even at the single particle level.
Exceptions include run-and-tumble particles in one dimension~\cite{Schnitzer1993,Tailleur2008,Tailleur2009,Malakar2018}, active Brownian particles (ABPs) in channels~\cite{Wagner2017} or sedimenting in a gravitational field~\cite{Hermann2018}, and ABPs confined in a harmonic well~\cite{Malakar2020,Caraglio2022}.
Finally, in free space, the characterization of the time-dependent probability distribution of run-and-tumble and ABPs is known only in the Fourier domain~\cite{Martens2012,Sevilla2015,Kurzthaler2016,Kurzthaler2017,Kurzthaler2018}.

In nature, several processes are naturally characterized in terms of the time required to reach for the first time a specific site (or a set of sites).
Examples include emission of an electrical impulse from a neuron~\cite{Gerstner1997}, execution of buy/sell orders in financial markets~\cite{Sazuka2009,Baldovin2015}, chemical processes such as fluorescence quenching~\cite{Chmeliov2013}, and molecules diffusing in the interior of the cell and being absorbed at the cell boundaries~\cite{Biswas2016}.
Hence it is important to compute quantities such as the first-passage time distribution and the survival probability~\cite{Redner2001,Metzler2013,Risken1989,Palyulin2019}.
First-passage probability, also plays a pivotal role in understanding transport properties and escape dynamics of nano- and micro-particles or living micro-organisms in different environments~\cite{Redner2001}.
However, while first-passage-time distribution of passive Brownian particles has been widely investigated~\cite{Klein1952,Kim2015,Chatterjee2018,Durang2019,Besga2021}, less is known about its properties in the case of self-propelled particles.
In fact, first-passage properties of active particles has been only recently investigated in a class of one-dimensional models~\cite{Malakar2018,Angelani2014,Scacchi2018,Demaerel2018,Dhar2019} and as the probability to hit a wall for a two-dimensional ABP without translational diffusion~\cite{Basu2018}.
In the latter case, ABP exhibits anomalous first-passage properties at short times which are a fingerprint of its non-equilibrium dynamics.

Here, we investigate the behavior of an ABP moving in a circular region with an absorbing boundary.
Taking inspiration from the approach used to solve the ABP model in a harmonic well~\cite{Caraglio2022}, we show that a formally exact series expression for the probability propagator can be obtained starting from the basis states of a reference standard Brownian particle model. 
Not only we provide the expression of the propagator, but by properly integrating the latter, we also find the survival probability and the first-passage-time distribution.

\section*{Model}

The stochastic overdamped motion of a two-dimensional ABP is completely characterized in terms of the propagator $\mathbb{P}(\vec{r}, \vartheta, t | \vec{r}_0 , \vartheta_0)$ which is the probability to find the particle at position $\vec{r}$ and orientation $\vartheta$ at lag time $t$ given the initial position $\vec{r}_0$ and orientation $\vartheta_0$ at time $t=0$.
In free space, the Fokker-Planck equation~\cite{Sevilla2015,Kurzthaler2018} reads
\begin{align} \label{eq:eom_propagator}
    \partial_t\mathbb{P}  = \mathcal{L} \mathbb{P} :=
    &  D \nabla^2 \mathbb{P} + D_{\text{rot}} \partial_\vartheta^2 \mathbb{P} -  v \vec{u} \cdot \vec{\nabla} \mathbb{P} \; ,
\end{align}
and readily provides the formal solution~\cite{Risken1989} 
\begin{align} \label{eq:formal_solution} \mathbb{P}(\vec{r}, \vartheta, t | \vec{r}_0 , \vartheta_0) = e^{\mathcal{L}t} \delta(\vec{r}-\vec{r}_0) \delta(\vartheta-\vartheta_0) \; ,
\end{align}
of the propagator given the initial condition
\begin{align} \label{eq:initial_condition}
    \mathbb{P}(\vec{r}, \vartheta, t=0 | \vec{r}_0 , \vartheta_0) =  \delta(\vec{r}-\vec{r}_0) \delta(\vartheta-\vartheta_0) \; .
\end{align}
Here, the first and the second term on the r.h.s. of Eq.~\eqref{eq:eom_propagator} respectively encode the translational and rotational diffusion of the particle, with diffusion coefficients $D$ and $D_{\text{rot}}$, respectively.
Finally, the last term corresponds to the self-propulsion of the particle having fixed modulus $v$ and directed along the orientation $\vec{u} = (\cos\vartheta, \sin \vartheta)$.

Here, we are interested in investigating the dynamics of an ABP conditioned to the presence of a circular absorbing boundary with radius $R$ and centered in the origin. 
This coincides with imposing the boundary condition
\begin{align} \label{eq:boundary_condition}
\mathbb{P}(\vec{R}, \vartheta, t | \vec{r}_0 , \vartheta_0) = 0 \; ,
\end{align}
and the further requirement that the initial position $\vec{r}_0$ is chosen inside the boundary, $r_0=|\vec{r}_0| < R$.
The radius $R$ fixes the length unit of the problem.
Taking the passive Brownian particle ($v=0$) as a reference, it is convenient to define the time unit $\tau$ as the typical time required by such a particle to reach the boundary, $\tau := R^2/D$.
The dynamics of the ABP is thus described only by two independent dimensionless parameters: The P\'eclet number, $\text{Pe} = v \tau /R$, assessing the importance of the self-propulsion with respect to the diffusive motion, and the \textit{``rotationality''}, $\gamma =  D_{\text{rot}} \tau$,  measuring the magnitude of the rotational diffusion.
The fraction $\text{Pe}/\gamma$ then describes the persistence of the particle's trajectories.

In order to find an expression for the propagator that is a solution of Eq.~\eqref{eq:eom_propagator}, first we make a time-separation ansatz for the propagator, $\mathbb{P}=\mathcal{E}(t)\psi(\vec{r},\vartheta)$~\cite{Risken1989}.
Inserting into~\eqref{eq:eom_propagator} and using the dimensionless parameters yields
\begin{equation}\label{eq:SE3}
    \frac{1}{\mathcal{E}}\partial_t \mathcal{E} = \frac{1}{\tau \psi} \left[ R^2 \nabla^2 \psi + \gamma \, \partial_\vartheta^2 \psi -\text{Pe} \, R \, \vec{u}\cdot\vec{\nabla} \psi \right]  \stackrel{!}{=} -\lambda \;,
\end{equation}
where the last equal sign holds since the first and second term of the equation are functions of independent variables, and therefore can only be equal to each other if their value is independent of all variables.
We can now explicitly write the solution for the time-dependent component of the propagator, $\mathcal{E}(t)=\exp(-\lambda t)$, and proceed to solve the equation for the spatial and angular components only, which now reads
\begin{equation}
    \mathcal{L}\psi + \lambda \psi = 0 \; .
\end{equation}
The similarity to a quantum mechanical problem suggests to tackle the problem in a perturbative approach. 
We thus split the Fokker-Planck operator $\mathcal{L}$ into an equilibrium contribution $\mathcal{L}_0$ and a non-equilibrium driving $\mathcal{L}_1$ according to
\begin{equation}
    \mathcal{L} = \mathcal{L}_0 + \text{Pe} \, \mathcal{L}_1 \; ,
\end{equation}
where, in polar coordinates $\vec{r} = r (\cos \varphi, \sin \varphi)$,
\begin{gather}
    \mathcal{L}_0 \psi= 
    \dfrac{1}{\tau} \left[ \frac{R^2}{r}\partial_r (r\partial_r\psi) + \frac{R^2}{r^2}\partial_\varphi^2\psi + \gamma \partial_\vartheta^2\psi \right] \;, \label{eq:L0} \\
    \mathcal{L}_1\psi=\dfrac{1}{\tau} \left[ -R\cos(\vartheta \!-\! \varphi)  \partial_r \psi - \frac{R}{r}\sin(\vartheta \!-\! \varphi)  \partial_\varphi \psi \right] \;. \label{eq:L1}
\end{gather}

\section*{Solution of the equilibrium reference system}
To solve the unperturbed eigenvalue problem 
\begin{equation} \label{eq:SE4}
    \mathcal{L}_0 \psi = - \lambda \psi \; ,
\end{equation}
subjected to the initial~\eqref{eq:initial_condition} and the boundary~\eqref{eq:boundary_condition} conditions, first we decompose the three degree of freedom in polar coordinates, $(r,\varphi,\vartheta)$ into different $(n,\ell,j)$ modes with the ansatz
\begin{equation} \label{eq:eigenfunctions_ansatz}
    \psi_{n,\ell,j}(\vec{r},\vartheta)=e^{i \ell \varphi} e^{i (j-\ell)\vartheta} \mathcal{R}_{n,\ell,j}(r) \; .
\end{equation}
This ansatz reflects the symmetries of the problem: 
Eq.~\eqref{eq:L0} shows that $\mathcal{L}_0$ is unchanged under a shift of the orientation angle $\vartheta$ or of the azimuthal angle $\varphi$ of the particle, meaning that it commutes with the corresponding generators $L=-i \partial_\varphi$ and $S=-i \partial_\vartheta$.
Borrowing a quantum mechanics language, we refer to these generators respectively as `orbital momentum' and `spin'.
Thus, the eigenfunctions $\psi_{n,\ell,j}$ of the passive reference system ($\text{Pe}=0$) are simultaneous eigenfunctions to orbital momentum and spin with eigenvalues $\ell$ and $s:= j-\ell$, respectively.
Here the spin $s$ is conveniently written in terms of $j$, i.e the eigenvalue of the total `angular momentum' $J = L + S$.
In fact, for the active particle ($\text{Pe}>0$) the Fokker-Planck operator $\mathcal{L}$ remains invariant only under a simultaneous rotation of position and orientation, such that in the full problem $j$ will be a conserved quantum number, simplifying future analytics and numerics.
Although the ansatz to solve the Fokker-Planck equation can often to be taken as real~\cite{Risken1989}, it is convenient to take a complex ansatz~\eqref{eq:eigenfunctions_ansatz}, in analogy to what done to solve various problem in electrodynamics~\cite{LandauLifshitzED}.

Inserting Eq.~\eqref{eq:eigenfunctions_ansatz} into~(\ref{eq:SE4}) yields the radial equation
\begin{align}
    r^2  \partial_r^2 \mathcal{R}_{n,\ell,j} & + r \, \partial_r \mathcal{R}_{n,\ell,j} -\ell^2 \, \mathcal{R}_{n,\ell,j} \nonumber \\ 
    + & \frac{r^2}{R^2} \left[  \lambda \tau - \gamma(j-\ell)^2 \right]  \mathcal{R}_{n,\ell,j}=0 \; ,
\end{align}
which, with the substitution $z^2= \frac{r^2}{R^2} [\lambda \tau-\gamma(j-\ell)^2] $, can be rewritten as
\begin{align}
    z^2 \partial_z^2 \mathcal{R}_{n,\ell,j} + z \partial_z \mathcal{R}_{n,\ell,j} + \left[ z^2-\ell^2 \right] \mathcal{R}_{n,\ell,j} = 0 \; . 
\end{align}
This is a Bessel equation of order $\ell$ with solutions $\mathcal{R}_{n,\ell,j}(z)=\text{J}_{\ell}(z) $, where $\text{J}_{\ell} (\cdot)$ is the Bessel function of the first kind of order $\ell$~\cite{NIST}.
Being the boundary condition~\eqref{eq:boundary_condition} valid at all times $t$, it must be satisfied by each eigenfunction $\psi_{n,\ell,j}$  independently.
This implies that $\sqrt{\lambda_{n,\ell,j}\tau-\gamma(j-\ell)^2}$ must be a root of the Bessel function $\text{J}_{\ell}(\cdot)$ for each value of $(n,\ell,j)$, which yields the discrete spectrum for the eigenvalues $\lambda_{n,\ell,j}$
\begin{equation} \label{eq:EVa}
    \lambda_{n,\ell,j}= \dfrac{1}{\tau} \left[ \text{j}_{\ell,n}^2 + \gamma (j-\ell)^2 \right] \;,
\end{equation}
where $\text{j}_{\ell,n}$ is the $n$-th root of the Bessel function $\text{J}_{\ell}$.

The explicit expression of the eigenfunctions reads
\begin{equation} \label{eq:eigenfunctions}
    \psi_{n,\ell,j}(\vec{r},\vartheta)=e^{i \ell \varphi} e^{i (j-\ell)\vartheta} \dfrac{\text{J}_{\ell} \! \left( \text{j}_{\ell,n} \dfrac{r}{R} \right)}{\sqrt{2} \pi R \text{J}_{\ell+1}(\text{j}_{\ell,n})}  \; ,
\end{equation}
where the normalization constant has been chosen such that
\begin{align} \label{eq:normalization}
    \braket{\psi_{n',\ell',j'}}{\psi_{n,\ell,j}} = \delta_{n,n'} \, \delta_{\ell,\ell'} \, \delta_{j,j'} \; .
\end{align}
Here we introduced the Kubo scalar product
\begin{align} \label{eq:KuboScalarProduct}
    \braket{\phi}{\psi} := \int_{r \leq R}  \diff \vec{r} \int_0^{2\pi} \diff \vartheta \, \phi(\vec{r},\vartheta)^* \psi(\vec{r},\vartheta) \; ,
\end{align}
and resorted on the relation~\cite{NIST}
\begin{equation}\label{eq:BOC}
    \int_0^R \diff r \, r \, \text{J}_{\ell} \! \left( \text{j}_{\ell,n} \, \frac{r}{R} \right)  \text{J}_{\ell}\! \left( \text{j}_{l,n'} \frac{r}{R} \right) = \frac{R^2}{2} \left[ \text{J}_{\ell+1}(\text{j}_{\ell,n}) \right]^2 \, \delta_{n,n'} \; .
\end{equation}
The isomorphism between $\ket{\psi}$ and $\psi(\vec{r},\vartheta)$ is made explicit by introducing generalized position and orientation states $|\vec{r} \vartheta\rangle$ such that $\psi(\vec{r},\vartheta) = \langle \vec{r}\vartheta |\psi \rangle$.
Using the orthogonality condition~\eqref{eq:normalization} it is easy to see that the operators $\ket{\psi_{n,\ell,j}} \bra{\psi_{n,\ell,j}}$ are a set of orthogonal projectors and thus we can write the following identity relation
\begin{align} \label{eq:completeness_Hilbert_space}
    \sum_{n,\ell,j} \ket{\psi_{n,\ell,j}} \bra{\psi_{n,\ell,j}} = \mathbb{1} \; ,
\end{align}
where we introduced a compact notation for the summation
\begin{align}
    \sum_{n,\ell,j}  :=  \sum_{n=1}^\infty \sum_{\ell=-\infty}^\infty \sum_{j=-\infty}^\infty  \; .
\end{align}
Note that the sum over the index $n$ is running only over strictly positive integer numbers because of the convention of counting nontrivial roots of the Bessel functions.
Finally, note also that, by making use of the completeness relation for Bessel function~\cite{NIST}
\begin{align} \label{eq:completeness_Bessel}
    \sum_{n=1}^{\infty} \dfrac{\text{J}_\ell (\text{j}_{\ell,n}z) \, \text{J}_\ell (\text{j}_{\ell,n}z_0)}{\left[ \text{J}_{\ell+1} (\text{j}_{\ell,n}) \right]^2} = \dfrac{1}{2z} \delta(z-z_0) \; ,
\end{align}
the eigenfunctions of the equilibrium reference system fulfill the completeness relation
\begin{align} \label{eq:completeness}
    \sum_{n,\ell,j} \psi_{n,\ell,j}(\vec{r},\vartheta) \psi_{n,\ell,j}(\vec{r}_0,\vartheta_0)^* =  \delta(\vec{r}-\vec{r}_0) \delta(\vartheta - \vartheta_0) \; .
\end{align}

The previous completeness relation~\eqref{eq:completeness} allows us to find a solution for the propagator in the equilibrium reference system starting from its formal expression~\eqref{eq:formal_solution}
\begin{align}\label{eq:solution_propagator}
    \mathbb{P}^{(0)} &(\vec{r}, \vartheta, t | \vec{r}_0 , \vartheta_0)  = \sum_{n,\ell,j} \left\lbrace e^{\mathcal{L}_0 t} \psi_{n,\ell,j}(\vec{r},\vartheta)  \right\rbrace \psi_{n,\ell,j}(\vec{r}_0,\vartheta_0)^* \nonumber \\ 
    & = \sum_{n,\ell,j} \bra{\vec{r}\vartheta} e^{\mathcal{L}_0 t} \ket{ \psi_{n,\ell,j}} \braket{\psi_{n,\ell,j}}{\vec{r}_0 \vartheta_0} \nonumber \\
    & = \sum_{n,\ell,j} e^{-\lambda_{n,\ell,j}t} \, \psi_{n,\ell,j}(\vec{r}_0 ,\vartheta_0)^* \, \psi_{n,\ell,j}(\vec{r} ,\vartheta)
\; .
\end{align}
Note that, from the second line of the previous equation and using the identity relation~\eqref{eq:completeness_Hilbert_space}, one can also write
\begin{align}\label{eq:solution_propagator_2}
    \mathbb{P}^{(0)} &(\vec{r}, \vartheta, t | \vec{r}_0 , \vartheta_0)  =  \bra{\vec{r}\vartheta} e^{\mathcal{L}_0 t} \ket{\vec{r}_0 \vartheta_0}
\; ,
\end{align}
meaning that the propagator is the projection of the generalized position and orientation state $\ket{\vec{r}\vartheta}$ over the time evolution of the initial state $\ket{\vec{r}_0\vartheta_0}$.

\section*{Solution for ABP particles}

One readily shows that the equilibrium operator $\mathcal{L}_0$ is Hermitian, $\braket{\phi}{\mathcal{L}_0\psi} = \braket{\mathcal{L}_0 \phi}{\psi}$, with respect to the Kubo scalar product~\eqref{eq:KuboScalarProduct} and consequently its eigenvalues $\lambda_{n,\ell,j}$ are real and left and right eigenfunctions  coincide, $\ket{\psi_{n,\ell,j}^{\text{L}}} = \ket{\psi_{n,\ell,j}^{\text{R}}} =\ket{\psi_{n,\ell,j}}$.
However, the full operator $\mathcal{L}$ does not reflect this property, in particular $\mathcal{L}_1$ is anti-Hermitian, $\braket{\phi}{\mathcal{L}_1\psi} = -\braket{\mathcal{L}_1 \phi}{\psi}$.
Correspondingly, in the following one has to be careful that the eigenvalues of the full operator, $\lambda_{n,\ell,j}^{\text{Pe}}$ are in general complex and the left eigenfunctions, $\ket{\psi_{n,\ell,j}^{\text{Pe},\text{L}}}$, are distinct from the right ones $\ket{\psi_{n,\ell,j}^{\text{Pe},\text{R}}}$.

If properly normalized, the perturbed left and right eigenfunctions constitute a bi-orthonormal basis with identity relation
\begin{align} \label{eq:completeness_Hilbert_space_perturbed}
    \sum_{n,\ell,j} \ket{\psi_{n,\ell,j}^{\text{Pe},\text{R}}} \bra{\psi_{n,\ell,j}^{\text{Pe},\text{L}}} = \mathbb{1} \; ,
\end{align}
which directly yields the propagator of the full problem
\begin{align}\label{eq:solution_propagator_full_problem}
    \mathbb{P} &(\vec{r}, \vartheta, t | \vec{r}_0 , \vartheta_0)  =  \bra{\vec{r}\vartheta} e^{\mathcal{L} t} \ket{\vec{r}_0 \vartheta_0} \nonumber \\
    & = \sum_{n,\ell,j} \bra{\vec{r}\vartheta} e^{\mathcal{L} t} \ket{ \psi_{n,\ell,j}^{\text{Pe},\text{R}}} \braket{\psi_{n,\ell,j}^{\text{Pe},\text{L}}}{\vec{r}_0 \vartheta_0} \nonumber \\
    & = \sum_{n,\ell,j} e^{-\lambda_{n,\ell,j}^{\text{Pe}}t} \, \psi_{n,\ell,j}^{\text{Pe},\text{L}}(\vec{r}_0 ,\vartheta_0)^* \, \psi_{n,\ell,j}^{\text{Pe},\text{R}}(\vec{r} ,\vartheta)
\; .
\end{align}

To explicitly compute the full propagator~\eqref{eq:solution_propagator_full_problem}, it is then necessary to calculate the perturbed eigenvalues and left and right eigenfunction.
To this scope, one has first to explicitly evaluate the action of the perturbation $\mathcal{L}_1$ on the eigenstates of $\mathcal{L}_0$.
Starting from Eqs.~\eqref{eq:L1} and~\eqref{eq:eigenfunctions} is it possible to show that
\begin{align}\label{eq:L1action}
    \mathcal{L}_1 \ket{\psi_{n,\ell,j}} = \sum_{n'=1}^{\infty} \left[ c^{+}_{n',n,\ell} \ket{\psi_{n',\ell+1,j}} + c^{-}_{n',n,\ell} \ket{\psi_{n',\ell-1,j}} \right] \; ,
\end{align}
with weights
\begin{align}\label{eq:L1action_coeffs}
    c^{\pm}_{n',n,\ell} = \dfrac{\text{j}_{\ell,n} \, \text{j}_{\ell \pm 1,n'}}{(\text{j}_{\ell,n}^2 - \text{j}_{\ell \pm 1,n'}^2)} \dfrac{\text{J}_{\ell \pm 1}(\text{j}_{\ell ,n}) \, \text{J}_{\ell}(\text{j}_{\ell \pm 1 ,n'})}{\text{J}_{\ell+1}(\text{j}_{\ell,n}) \, \text{J}_{\ell+1 \pm 1}(\text{j}_{\ell \pm 1,n'})} \; ,
\end{align}
(see \hyperref[sec_appA]{appendix A} for further details).

Now, considering the finite-dimensional subspace of equilibrium eigenfunctions such that $0<n\leq n_{\text{max}}$, $|\ell| \leq \ell_{\text{max}}$, and $|j| \leq j_{\text{max}}$, the action of $\mathcal{L}=\mathcal{L}_0 + \text{Pe} \,\mathcal{L}_1$ is completely characterized by a square matrix of dimension $n_{\text{max}} (2\ell_{\text{max}} + 1)(2j_{\text{max}} + 1)$, which has to be diagonalized numerically to obtain its eigenvalues $\lambda_{n,\ell,j}^{\text{Pe}}$ and left and right eigenvectors, $\bra{\psi_{n,\ell,j}^{\text{Pe},\text{L}}}$ and $\ket{\psi_{n,\ell,j}^{\text{Pe},\text{R}}}$ for any given P\'eclet number.
Since the perturbation doesn't couple states with $j'\neq j$, the action of $\mathcal{L}$ is more conveniently characterized by defining different matrices $A_j$ for different $j$ channels, each with dimension $n_{\text{max}} (2\ell_{\text{max}} + 1)$ and elements defined by
\begin{eqnarray}
[A_j]_{n'-1+n_{\text{max}} ( \ell' + \ell_{\text{max}}) ,  n-1+n_{\text{max}} ( \ell + \ell_{\text{max}})} \nonumber \\
\quad = \bra{\psi_{n',\ell',j}} \left(  \mathcal{L}_0 + \text{Pe} \, \mathcal{L}_1 \right)  \ket{\psi_{n,\ell,j}} \; .
\end{eqnarray}
The perturbed eigenvectors are then a linear combination of the equilibrium eigenstates
\begin{align}
\ket{\psi_{n,\ell,j}^{\text{Pe},\text{R}}} & = \sum_{n',\ell'} a_{n,\ell,j}^{\text{R},n',\ell'} \ket{\psi_{n',\ell',j}} \; , 
\label{eq:decomposition_eigenstates1} \\
\bra{\psi_{n,\ell,j}^{\text{Pe},\text{L}}} & = \sum_{n',\ell'} a_{n,\ell,j}^{\text{L},n',\ell'} \bra{\psi_{n',\ell',j}} \; . \label{eq:decomposition_eigenstates2}
\end{align}

\begin{figure}[t!]
\centering
\includegraphics[scale=1]{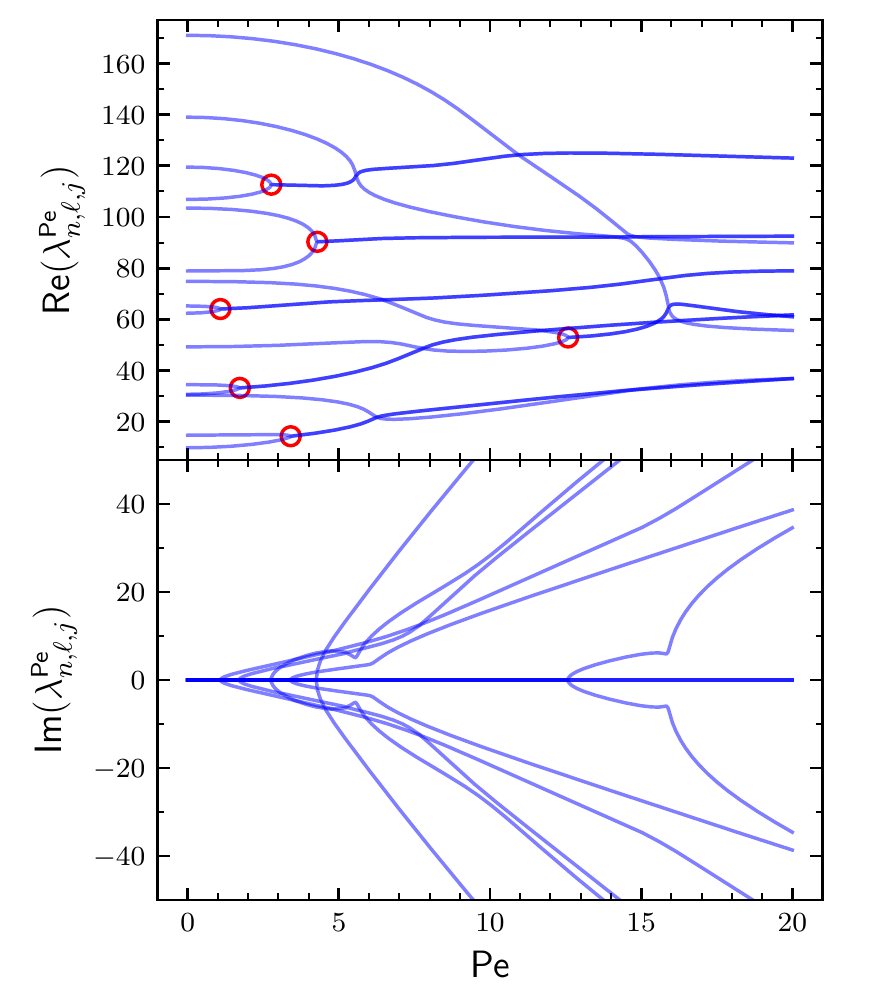} 
\caption{Numerical eigenvalues $\lambda_{n,\ell,j}$ of the Fokker-Planck operator $\mathcal{L} = \mathcal{L}_0+\text{Pe} \, \mathcal{L}_1$ as a function of the P{\'e}clet number $\text{Pe}$, for $\gamma=4$, $n_{\text{max}}=3$, $\ell_{\text{max}}=2$, and $j=1$. Transparency of lines and exceptional points highlighted with red circles better show when real components merge and imaginary ones bifurcate.  \label{fig:spectrum}}
\end{figure}

The computational time required to diagonalize these matrices increases rapidly with the dimension of the considered subspace. 
However, the decaying exponentials in time in the expression of the propagator, Eq.~\eqref{eq:solution_propagator_full_problem}, ensures convergence.
In the unperturbed case, $\text{Pe}=0$, the eigenvalues~\eqref{eq:EVa} are real and an increasing function of $n$, $|\ell|$ and $|j-\ell|$.
With increasing activity, more and more eigenvalues become complex quantities, and their original order in the real plane may change, see Fig.~\ref{fig:spectrum}.
Consequently, the numerical cutoffs $n_{\text{max}}$ and $\ell_{\text{max}}$ must be chosen accordingly.
Figure~\ref{fig:spectrum} suggests that for small enough value of the activity, for which the finite cut-off spectrum remains real, a perturbative approach adopting only real eigenfunctions would also solve the problem.
Interestingly enough, eigenvalues move to the complex plane in pairs: With increasing P{\'e}clet number, two real eigenvalues merge and bifurcate to a pair of complex conjugates for even larger activity. 
These branching points, called exceptional points~\cite{Heiss2012}, often originate in parameter-dependent eigenvalue problems and occur in a great variety of physical problems including mechanics, electromagnetism, atomic and molecular physics, quantum phase transitions, and quantum chaos. They even occur in other problems concerning active particles~\cite{Kurzthaler2016,Kurzthaler2017}.
The exceptional points are highlighted with red circles in the upper panel of Fig.~\ref{fig:spectrum}.
Note also that, in line with the non-crossing rule between eigenvalues of an Hermitian matrix representing a quantum observable~\cite{LandauLifshitzQM}, crossings between the real components of the eigenvalues occur only when at least one of the eigenvalues participating to the crossing has already become a complex quantity.

\begin{figure*}[t!]
\centering
\includegraphics[scale=1]{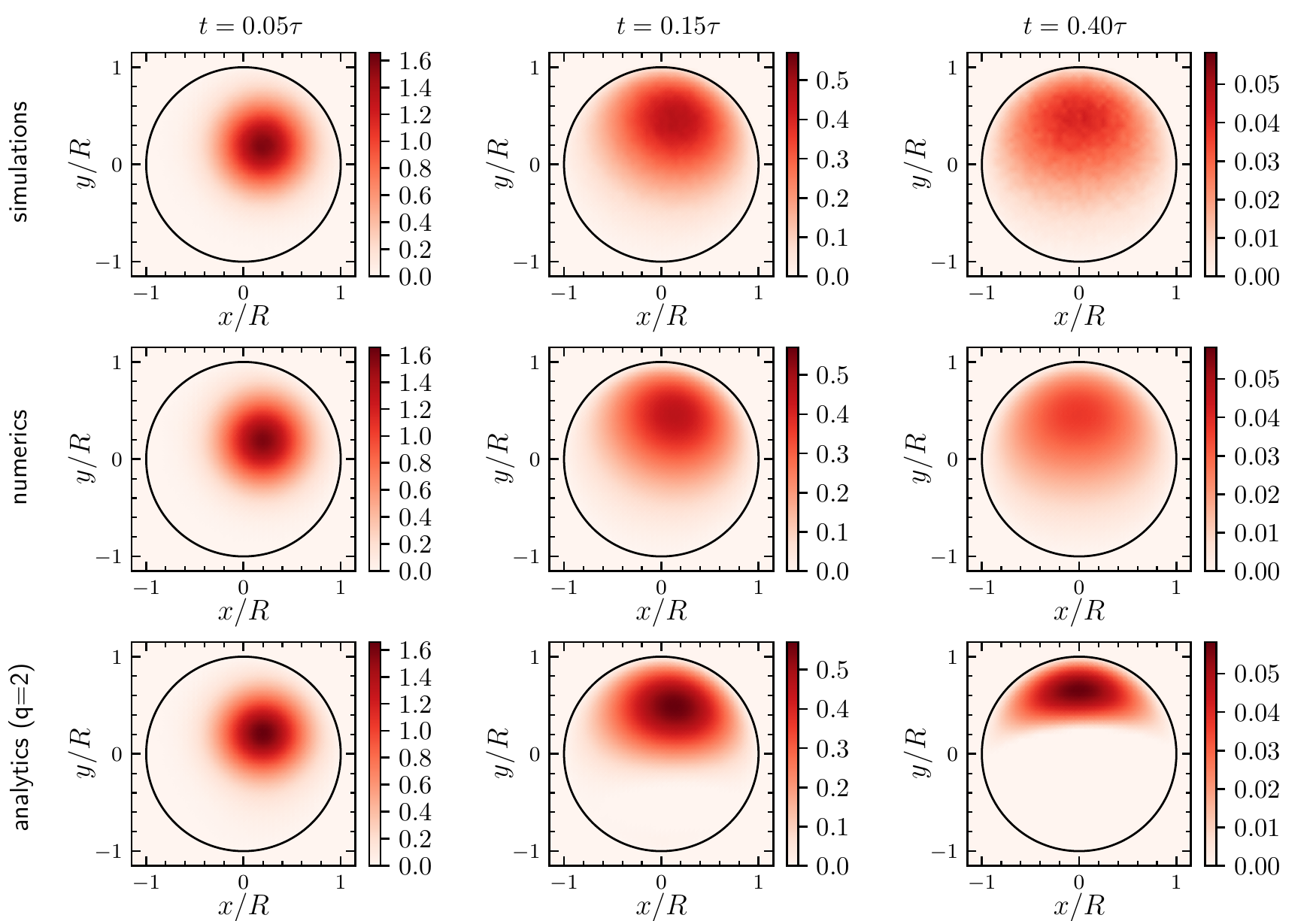}
\caption{Spatial probability distribution at different times $t$ starting with initial condition $r_0 = 0.2R$, $\varphi_0 = 0$, and $\vartheta_0 = \pi/2$. Comparison between simulations, numerics, and analytics up to second order perturbation theory ($q=2$) for $\text{Pe} = 4$ and $\gamma=0.8$.
For the simulations, statistics has been collected from $10^7$ independent particles.
For the numerics and the analytics, $n_{\text{max}}=8$, $\ell_{\text{max}}=j_{\text{max}}=7$.\label{fig:spatial_probability}}
\end{figure*}

\begin{figure*}[t!]
\centering
\includegraphics[scale=1]{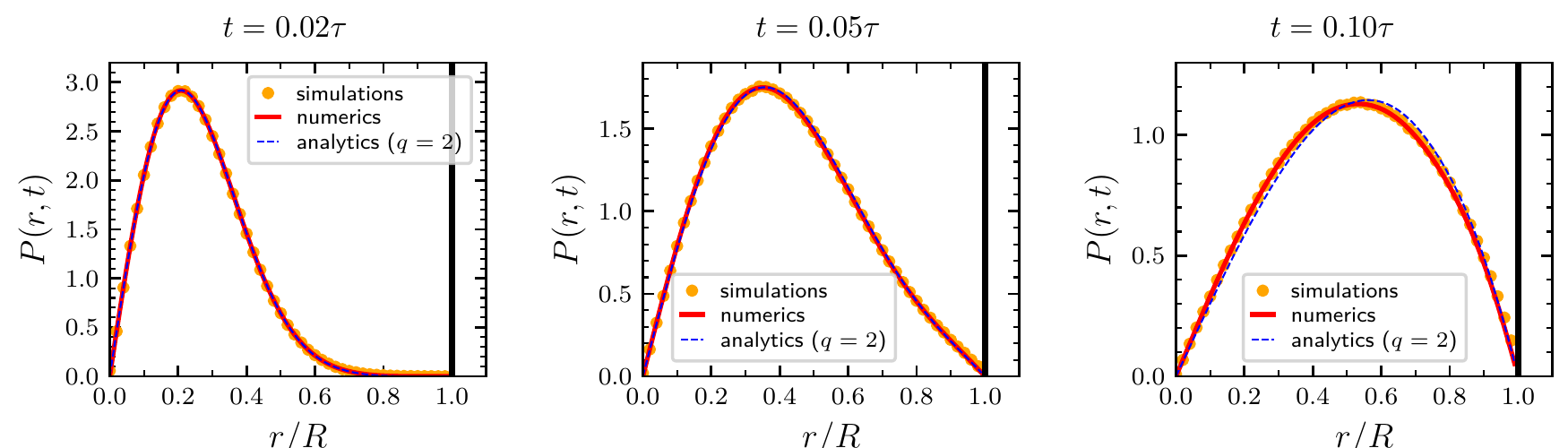}
\caption{Radial probability distribution at different times $t$ starting with initial condition $r_0 = 0$, averaged over $\vartheta_0$. 
Normalization is such that 
$\int_0^R P(r,t) \, \diff r = S(t)$, with $S(t)$ the survival probability of the particle at time $t$.
Comparison between simulations, numerics, and analytics up to second order perturbation theory ($q=2$).
For the simulations, statistics has been collected from $10^7$ independent particles.
For the numerics, $n_{\text{max}}=8$, $\ell_{\text{max}}=j_{\text{max}}=7$.\label{fig:radial_probability}}
\end{figure*}

To corroborate our findings, we benchmark the time evolution of the spatial probability distribution starting from some given initial condition as obtained from numerics against that obtained by direct stochastic simulations, see Fig.~\ref{fig:spatial_probability}.
The radial probability density, given the initial position $\vec{r}_0$ and averaged over a uniform distribution of the initial direction $\vartheta_0$, defined by
\begin{align}
P(r,t | \vec{r}_0) \!=\! r \! \int_0^{2\pi} \!\! \dfrac{\diff \vartheta_0}{2\pi} \! \int_0^{2\pi} \!\!\!\! \diff \vartheta \! \int_0^{2\pi} \!\!\!\! \diff \varphi \, \mathbb{P}(\vec{r},\vartheta,t|\vec{r_0},\vartheta_0) ,
\end{align}
is reported in Fig.~\ref{fig:radial_probability}.
Note that in the previous expression, normalization has been chosen such that
$\int_0^R \diff r P(r,t | \vec{r}_0)$ represents the probability that at time $t$ the particle has not yet reached the absorbing boundary at $r=R$.

A remark on stochastic simulations is in order: due to the presence of the absorbing boundary, the discretization time step should be smaller than what is usually adopted for standard simulations of an ABP in free space.
In particular, with increasing time, convergence of results in the proximity of the boundary becomes more and more sensitive to the value of the discretization time step.
See \hyperref[sec_appB]{appendix B} for more details.

\subsection*{Recursive analytical solution}

It is instructive to introduce a second approach that, following the framework of perturbation theory in quantum mechanics, allows obtaining the propagator as a combination of the unperturbed eigenfunctions weighted by factors that depend only on time and the initial conditions.
These functions consist of a power series in the P{\'e}clet number where the coefficient of the $q$-th order term can be computed in an exact iterative scheme.
In order to obtain these recursive relations, we start by rewriting the propagator~\eqref{eq:solution_propagator_full_problem} as
\begin{align}\label{eq:solution_propagator_full_problem_2}
    \mathbb{P} &(\vec{r}, \vartheta, t | \vec{r}_0 , \vartheta_0) = \sum_{n,\ell,j} M_{n,\ell,j}(\vec{r}_0,\vartheta_0,t) \, \psi_{n,\ell,j}(\vec{r} ,\vartheta)
\; ,
\end{align}
where
\begin{align}\label{eq:Ms}
M_{n,\ell,j}(\vec{r}_0,\vartheta_0,t) := \bra{\psi_{n,\ell,j}} e^{\mathcal{L}t} \ket{\vec{r}_0 \vartheta_0} \; .
\end{align}
Inserting the previous expression into the Dyson equation, familiar from quantum theory~\cite{SakuraiQM}, for the time evolution operator
\begin{align} \label{eq:Dyson_series}
e^{\mathcal{L} t} = e^{\mathcal{L}_0 t} + \text{Pe} \int_0^t \diff s \, e^{\mathcal{L}_0 (t-s)} \, \mathcal{L}_1 \, e^{\mathcal{L} s} \; ,
\end{align}
one obtains a useful integral relation for the functions $M$ appearing in the propagator
\begin{align} \label{eq:recursion_Ms}
& M_{n,\ell,j} (\vec{r}_0,\vartheta_0,t) =  \; e^{-\lambda_{n,\ell,j} t} \braket{\psi_{n,\ell,j}}{\vec{r}_0\vartheta_0} + \nonumber \\
& \qquad + \! \text{Pe} \! \int_0^t \!\! \diff s  \, \bigg[ e^{- \! \lambda_{n,\ell,j} (t-s)} \!\! \nonumber \\
& \times \sum_{n',\ell',j'} \! \bra{\psi_{n,\ell,j}} \! \mathcal{L}_1 \! \ket{\psi_{n',\ell',j'}} 
M_{n',\ell',j'}(\vec{r}_0,\! \vartheta_0,\!s) \bigg] \, .
\end{align}
Such a relation may be explicitly calculated since the action of the perturbation $\mathcal{L}_1$ on the equilibrium eigenfunctions is known, see Eq.~\eqref{eq:L1action}.
In particular, if we make a power-series expansion in $\text{Pe}$,
\begin{align}
M_{n,\ell,j} (\vec{r}_0,\vartheta_0,t) = \sum_{q=0}^{\infty} \text{Pe}^q M_{n,\ell,j}^{(q)} (\vec{r}_0,\vartheta_0,t)\; ,
\end{align}
Eq.~\eqref{eq:recursion_Ms} allows calculating the $q$-th order contribution once the $(q-1)$-th terms have been evaluated
\begin{align} \label{eq:recursion_Mqs}
& M_{n,\ell,j}^{(q)} (\vec{r}_0,\vartheta_0,t) = \int_0^t \!\! \diff s  \,  e^{- \! \lambda_{n,\ell,j} (t-s)} \!\! \nonumber \\
& \!\! \times \!\!  \sum_{n'} \! \bigg[ c^+_{n,n'\!,\ell\!-\!1}  M_{n'\!,\ell\!-\!1,j}^{(q-1)}(\vec{r}_0,\! \vartheta_0,\!s\!) \!+\! c^-_{n,n'\!,\ell\!+\!1} M_{n'\!,\ell\!+\!1,j}^{(q-1)}\!(\vec{r}_0,\! \vartheta_0,\!s\!)\bigg] ,
\end{align}
starting with the $0$-th order given by
\begin{align}
M_{n,\ell,j}^{(0)} (\vec{r}_0,\vartheta_0,t) = e^{-\lambda_{n,\ell,j}t} \psi_{n,\ell,j}(\vec{r}_0,\vartheta_0)^* \; .
\end{align}
The previous scheme is particularly efficient to compute the $M$ functions at the first few orders but this evaluation becomes quickly tedious with increasing order $q$.
An integrated version of Eq.~\eqref{eq:recursion_Mqs} is reported in \hyperref[sec_appC]{appendix C} for $q=1$ and $q=2$.
The spatial and the radial probability density obtained from implementing the previous scheme up to the second order are reported in Figs.~\ref{fig:spatial_probability} and~~\ref{fig:radial_probability}, respectively.
Some differences with respect to the results obtained by numerical simulation are observed.
In fact, the analytical solution truncated at a given order $q$ represents a situation which is neither a truly active particle (retrieved in the limit of $q\rightarrow \infty$) nor a passive Brownian particle ($q=0$).
In contrast to intuition, in general, it is not even a situation located somewhere in between these two limiting cases because the series in Eq.~\eqref{eq:solution_propagator_full_problem_2} presents contribution terms of both signs, with the signs depending on a complex interplay between the values of the quantum numbers $(n,\ell,j)$ and of the order $q$ and the time $t$ at which the $M$'s coefficients are computed.
Furthermore, the longer the time $t$, the higher should be the order $q$ and the cutoffs $n_{\text{max}}$, $\ell_{\text{max}}$, and $j_{\text{max}}$.
This is because the propagator contains terms proportional to all exponentials $\exp(-\lambda_{n,\ell,j} t)$, and in the long-time regime, $t \rightarrow \infty$, for each channel $j$, the dominant contribution comes from the lowest eigenvalue $\lambda_{n^\star,\ell^\star,j}$.
However, each coefficient $M_{n,\ell,j}(\vec{r}_0,\vartheta_0,t)$ contains terms proportional to $\exp(-\lambda_{n^\star,\ell^\star,j}t)$, provided that the order at which it is calculated is large enough.

\section*{Survival probability and first-passage-time distribution}

Once the propagator is known, the survival probability at time $t$, given some initial conditions $(\vec{r}_0, \vartheta_0)$, is readily obtained by integrating over the final position and orientation
\begin{align}\label{eq:survival_probability}
S(t|\vec{r}_0, \vartheta_0) \! = \! \int_0^R \!\!\! r \, \diff r \int_0^{2\pi} \!\!\! \diff \vartheta  \int_0^{2\pi} \!\!\! \diff \varphi \, \mathbb{P}(\vec{r},\vartheta,t|\vec{r_0},\vartheta_0) \; .
\end{align}
Since 
\begin{align}
\int_0^R \!\!\!\! r \, \diff r \! \int_0^{2\pi} \!\!\!\! \diff \vartheta \! \int_0^{2\pi} \!\!\! \diff \varphi \, \psi_{n,l,j}(\vec{r},\vartheta) = 2 \sqrt{2} \pi R \dfrac{\delta_{j,0} \delta_{\ell,0}}{\text{j}_{0,n}} \; ,
\end{align}
and using Eqs~\eqref{eq:solution_propagator_full_problem},~\eqref{eq:decomposition_eigenstates1} and~\eqref{eq:decomposition_eigenstates2}, one obtains
\begin{align}\label{eq:survival_probability2}
S(t|\vec{r}_0, & \vartheta_0) = 2 \sqrt{2} \pi R 
\sum_{n,\ell} e^{-\lambda_{n,\ell,0}^{\text{Pe}}t} \nonumber \\
& \times
\sum_{n',\ell'} a_{n,\ell,0}^{\text{L},n',\ell'} \psi_{n',\ell',0}(\vec{r}_0,\vartheta_0) 
\sum_{n''} \dfrac{a_{n,\ell,0}^{\text{R},n'',0}}{\text{j}_{0,n}}
 \; .
\end{align}
Alternatively, starting from Eq.~\eqref{eq:solution_propagator_full_problem_2}, the survival also reads
\begin{align}\label{eq:survival_probability3}
S(t|\vec{r}_0, & \vartheta_0) = 2 \sqrt{2} \pi R 
\sum_{n} \dfrac{M_{n,0,0}(\vec{r}_0,\vartheta_0,t)}{\text{j}_{0,n}} \; . 
\end{align}

\begin{figure}[t!]
\centering
\includegraphics[scale=1]{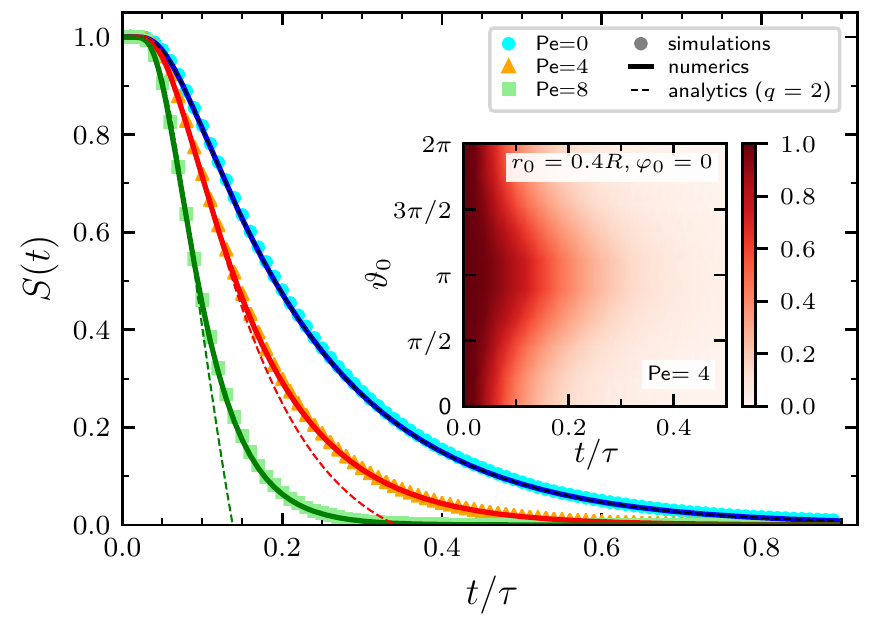}
\caption{Survival probability, $S(t) = S(t| \vec{r}_0,\vartheta_0)$, as a function of time for $\text{Pe} = 4$ and $\gamma=0.8$ and with initial condition $r_0=0$.
Comparison between simulations, numerics, and analytics ($q=2$).
For the simulations, statistics has been collected from $10^5$ independent particles.
For the numerics, $n_{\text{max}}=8$, $\ell_{\text{max}}=j_{\text{max}}=7$.
Inset: Survival probability as a function of time and $\vartheta_0$ as obtained from numerics for remaining initial conditions $r_0 = 0.4R$ and $\varphi_0 = 0$.
\label{fig:survival_probability}}
\end{figure}

See Fig.~\ref{fig:survival_probability} for a comparison between the results at different P{\'e}clet numbers obtained by numerics and by direct stochastic simulations.
As expected, due to their activity and the persistence of their motion, active particles tend to meet the boundary and be absorbed more often than passive particles, thus leading to a survival probability which is decaying faster.
Clearly, when $r_0 \neq 0$ the survival probability depends on the relative angle between the initial azimuth, $\varphi_0$, and the initial direction of the self-propulsion, $\vartheta_0$, and shows a slower decay when $\vartheta_0 = \pi + \varphi_0$, see inset of Fig.~\ref{fig:survival_probability}.

\begin{figure}[t!]
\centering
\includegraphics[scale=1]{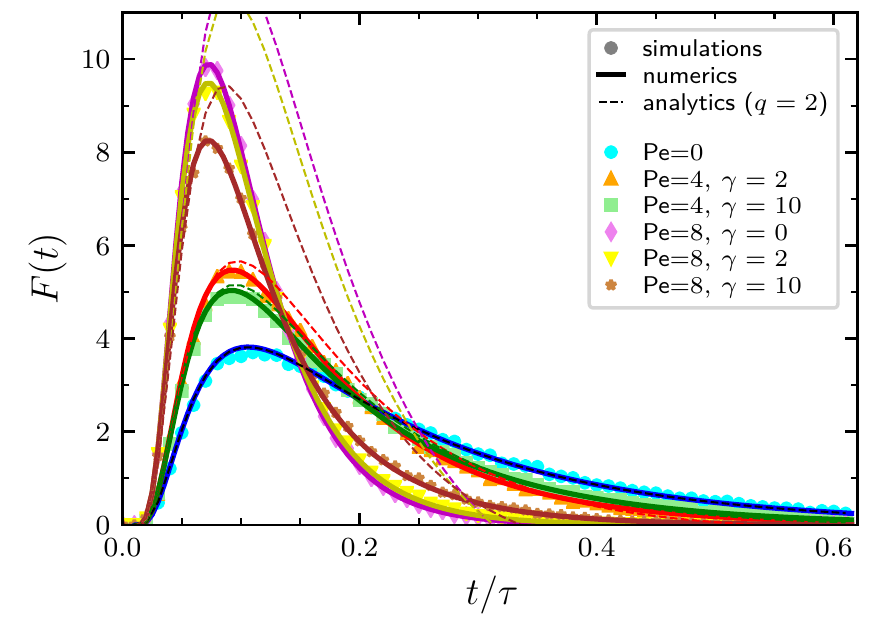}
\caption{First-passage-time distribution, $F(t) = F(t| \vec{r}_0,\vartheta_0)$, with initial condition $r_0=0$.
Comparison between simulations and numerics.
For the simulations, statistics has been collected from $10^7$ independent particles.
For the numerics, $n_{\text{max}}=10$, $\ell_{\text{max}}=j_{\text{max}}=9$.
\label{fig:fpt}}
\end{figure}

Finally, we can also compute the first-passage-time distribution for a given initial condition as
\begin{align}\label{eq:first_passage_def}
F(t|\vec{r}_0, \vartheta_0) = - \dfrac{\diff S(t|\vec{r}_0, \vartheta_0)}{\diff t} \; .
\end{align}
Such a distribution becomes sharper with increasing activity and decreasing rotational diffusivity.
However, the rotational diffusivity influences the first-passage-time distribution to an extent that also depends on the activity, see Fig.~\ref{fig:fpt}.
Figures~\ref{fig:survival_probability} and~\ref{fig:fpt} also report results obtained from the recursive analytical solution up to second order in perturbation, see dashed lines. As expected, while in the passive case these results coincide with those obtained by numerics, their quality degrades with increasing P{\'e}clet number.

\section*{Conclusions}

We have derived an exact series solution for the probability propagator of a two-dimensional ABP living in a circular region with absorbing boundary.
Such a solution is obtained by adopting standard passive Brownian motion as a reference system and dealing with the activity of the particle in a perturbative approach.
The propagator is then expressed in terms
of the left and right eigenvectors, which can be easily computed by direct diagonalization of the matrix form of
the Fokker-Planck operator, multiplied by an exponentially decaying factor with a rate given by the corresponding perturbed eigenvalue.
Using Dyson's approach, we also show that the propagator can be alternatively expressed as a combination of the unperturbed eigenfunctions weighted by
factors that depend only on time and the initial conditions and that can computed in an exact iterative scheme.
In principle, other geometries of the boundary may be considered using a similar framework. 
However, the choice of a circular domain implies that the total angular momentum is conserved, thus simplifying the analytics and the numerics.

Integration of the propagator over the coordinates at time $t$ allows obtaining the spatial probability density (integration over the self-propulsion orientation $\vartheta$), the radial probability distribution (integration over $\vartheta$ and the azimuth $\varphi$), and the survival probability (integration over $\vartheta$, $\varphi$, and the radius $r$).
Derivation of the latter with respect to time directly provides also the first-passage-time distribution, which displays a strong dependence on the activity of the particle and, to a lesser extent, on its rotational diffusivity.

Our results may be exploited to calculate other relevant observables such as the mean first-passage time and other moments of the first-passage time distribution~\cite{Redner2001}.
Furthermore, they can be generalized to chiral ABP~\cite{vanTeeffelen2008} by adding a drift term to the dynamics of the orientation of the particle or to a dynamics with resetting events~\cite{Chatterjee2018,Durang2019,Santra2020} by including in the model a probability for instantaneously returning to the initial condition after a certain time.
Finally, our findings may also serve as a starting point to investigate target-search~\cite{Tejedor2012,Volpe2017,Zanovello2021,Zanovello2021b} problems in complex environments involving absorbing boundaries.
However, it must be mentioned that generalizing our approach to account for one of the most interesting aspects of ABPs' dynamics, namely their accumulation in the presence of rigid boundaries~\cite{Bechinger2016}, remains an extremely challenging problem hindered by the fact that in these cases the evolution of the particle position is coupled to that of the self-propulsion direction in a way that depends on the position itself.

The approach allowing to derive our solution takes inspiration from the one recently adopted to solve the ABP model in a harmonic trap~\cite{Caraglio2022} and, in principle, can be adopted to solve other problems in which it is possible to solve the eigenvalue problem of the reference passive system.
Possible topics worth further investigation include escape rate kinetics~\cite{Goychuk2007} and statistics of the path times for particles crossing an energy barrier~\cite{Caraglio2018,Caraglio2020}.
However, it is worth mentioning that, in the current problem, as it often happens in parameter-dependent eigenvalues problem, changing the activity of the particle also induces a change in the eigenvalue spectrum of the Fokker-Planck operator, with exceptional points~\cite{Heiss2012} arising.
According to Eq.~\eqref{eq:solution_propagator_full_problem}, crossing these points by increasing the P\'eclet number implies that the propagator develops oscillating components because of the non-zero imaginary parts of the eigenvalues sharing the exceptional point.
In contrast, in the case of the harmonic oscillator, the Fokker-Planck operator of the passive Brownian particle and of the ABP are isospectral.
Since both problems are invariant under simultaneous rotations of the self-propulsion direction and the azimuthal angle, the isospectral property of the harmonic oscillator is surprising and induces to think that the harmonic potential problem should have some extra hidden symmetry, see the last remark in Ref.~\cite{Caraglio2022}.

\begin{acknowledgments}
T.F. acknowledges funding by FWF: P 35580-N.~~M.C. is supported by FWF: P 35872-N.
\end{acknowledgments}
\medskip

\appendix

\section{Action of $\mathcal{L}_1$ on the equilibrium eigenfunction $\ket{\psi_{n,\ell,j}}$} \label{sec_appA}
Here, we now explicitly evaluate the action of the perturbation $\mathcal{L}_1$ on the eigenstates of $\mathcal{L}_0$ obtained from the eigenvalue problem
\begin{align}
    \mathcal{L}_0 \ket{\psi_{n,\ell,j}} = -\lambda_{n,\ell,j} \ket{\psi_{n,\ell,j}} \; .
\end{align}
From the explicit expressions of the perturbation $\mathcal{L}_1$ and of the eigenfunction $\psi_{n,\ell,j}(\vec{r},\vartheta)$, Eq.s~\eqref{eq:L1} and~\eqref{eq:eigenfunctions}, and using the following properties of the Bessel functions
\begin{align}
    \frac{\diff}{\diff z} \text{J}_{\ell} (z) = \dfrac{\text{J}_{\ell-1} (z)-\text{J}_{\ell+1} (z)}{2} \; , \\
    \dfrac{\ell}{z} \text{J}_{\ell} (z) = \dfrac{\text{J}_{\ell-1} (z)+\text{J}_{\ell+1} (z)}{2} \; ,
\end{align}
one readily obtains
\begin{align} \label{eq:eq_appendix}
    \mathcal{L}_1 \psi_{n,\ell,j}& (\vec{r},\vartheta) = \dfrac{1}{\tau} \frac{\text{j}_{\ell,n}}{2\sqrt{2}\pi R\text{J}_{\ell+1}(\text{j}_{\ell,n})} \nonumber \\
    \times \Big\{ & e^{i(\ell+1)\varphi} e^{i(j-\ell-1) \vartheta}\, \text{J}_{\ell+1} \left( \text{j}_{\ell,n}\frac{r}{R} \right) \nonumber \\
    -& e^{i(\ell-1)\varphi} e^{i(j-\ell+1)\vartheta} \, \text{J}_{\ell-1} \left( \text{j}_{\ell,n}\frac{r}{R} \right) \Big\} \;. 
\end{align}
In order to write this expression as a linear superposition of eigenstates of $\mathcal{L}_0$, we use the completeness relation, Eq.~\eqref{eq:completeness_Bessel}, and Lommel's integral relation~\cite{NIST}
\begin{align}
    \int_0^1 \diff z \, z & \, \text{J}_{\ell \pm 1}(\text{j}_{\ell \pm 1, m} z)\, \text{J}_{\ell \pm 1}(\text{j}_{\ell, n} z)  = \nonumber \\
     & = \pm \frac{\text{j}_{\ell \pm 1, m} \, \text{J}_{\ell \pm 1}(\text{j}_{\ell, n}) \, \text{J}_{\ell}(\text{j}_{\ell \pm 1, m})}{\text{j}_{\ell, n}^2-\text{j}_{\ell \pm 1, m}^2} \; ,
\end{align}
to obtain
\begin{align}
    & \text{J}_{\ell \pm 1} \left( \text{j}_{\ell,n}\frac{r}{R} \right) = \int_0^1 \diff \! \left( \dfrac{r_0}{R} \right) \text{J}_{\ell \pm 1} \! \left( \text{j}_{\ell,n}\frac{r_0}{R} \right) \delta \! \left(\frac{r_0}{R}-\frac{r_0}{R} \right) \nonumber \\
    & \quad = 2 \sum_{m=1}^{\infty} \dfrac{\text{J}_{\ell \pm 1} \! \left( \text{j}_{\ell \pm 1,m} \frac{r}{R} \right)}{\left[ \text{J}_{\ell\pm 1 +1} (\text{j}_{\ell \pm 1,m}) \right]^2} \nonumber \\
    & \qquad \times \int_0^1 \diff \! \left( \dfrac{r_0}{R} \right) \left( \dfrac{r_0}{R} \right) \, \text{J}_{\ell \pm 1} \! \left( \text{j}_{\ell \pm 1,m} \frac{r}{R} \right) \, \text{J}_{\ell \pm 1} \! \left( \text{j}_{\ell,n}\frac{r_0}{R} \right) \nonumber \\
    & = \!\! \pm 2 \!\! \sum_{m=1}^{\infty} \! \frac{\text{j}_{\ell \pm 1, m} \, \text{J}_{\ell \pm 1}(\text{j}_{\ell, n}) \, \text{J}_{\ell}(\text{j}_{\ell \pm 1, m})}{(\text{j}_{\ell, n}^2 \!\!-\! \text{j}_{\ell \pm 1, m}^2) \! \left[ \text{J}_{\ell\pm 1 +1} (\text{j}_{\ell \pm 1,m}) \right]^2} \text{J}_{\ell \pm 1} \! \left( \text{j}_{\ell \pm 1,m} \frac{r}{R} \right) .
\end{align}
Finally, collecting the previous equation with Eq.~\eqref{eq:eq_appendix} we recover the action of the perturbation given in Eqs.~\eqref{eq:L1action} and~\eqref{eq:L1action_coeffs}.

\section{Dependence on discretization time step of stochastic simulations} \label{sec_appB}

\begin{figure}[t!]
\centering
\includegraphics[scale=1]{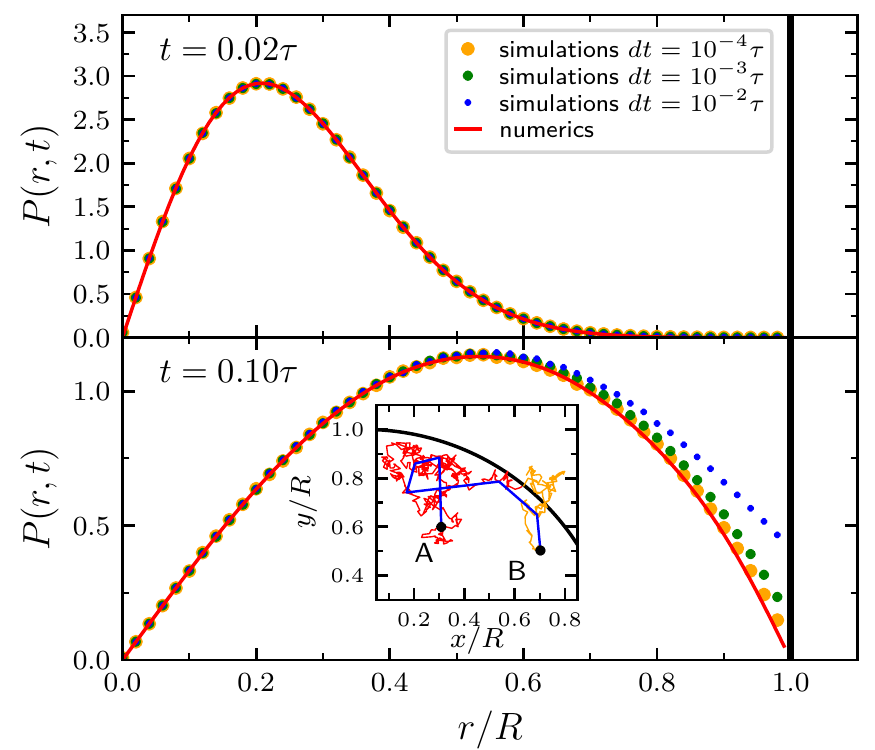}
\caption{Radial probability distribution (for $\text{Pe} = 4$ and $\gamma=0.8$, with initial condition $r_0 = 0$, and averaged over $\vartheta_0$) as obtained from numerics and from stochastic simulations assuming different values of $dt$. 
Upper panel $t=0.02\tau$, lower panel $t=0.1\tau$.
Inset: examples of two trajectories (Red $dt=10^{-4}\tau$, Blue $dt=10^{-2}\tau$) connecting the same points A and B in the proximity of the boundary.  \label{fig:dt_dependence}}
\end{figure}

Due to the presence of the absorbing boundary, the simulation results are more sensitive to the discretization time step than standard simulations in free space,
particularly in the proximity of the boundary, see Fig.~\ref{fig:dt_dependence} for an overview of results obtained for the radial probability density at different values of $dt$.

The reason for this sensitiveness lies in the fact that, given two points that are both close to the boundary but still inside the disk, a fine-grained trajectory connecting these two points has more chances to cross the boundary, thus not counting in the overall statistics, than a more coarse-grained trajectory.

\section{Numerical computation of the functions $M^{(q)}_{n,\ell,j}$} \label{sec_appC}

To compute numerically the coefficients $M^{(q)}_{n,\ell,j}(\vec{r}_0,\vartheta_0,t)$, having an integrated version of Eq.~\eqref{eq:recursion_Mqs} is convenient.
Unfortunately, eventual degeneracies in the eigenvalue spectrum cause difficulties in deriving an integrated recursive scheme valid at each order.
Here, we give expression for $q=1$ and $q=2$ as obtained by Eq.~\eqref{eq:recursion_Mqs} starting from the zero-th order
\begin{align}
M_{n,\ell,j}^{(0)} (\vec{r}_0,\vartheta_0,t) = e^{-\lambda_{n,\ell,j}t} \psi_{n,\ell,j}(\vec{r}_0,\vartheta_0)^* \; .
\end{align}
For the sake of compactness, in the rest of this section, we suppress the dependence on the initial conditions of the $M$ coefficients.
Introducing the characteristic function
\begin{align}
\chi_{n,\ell,j}^{n',\ell',j'} = \left\lbrace 
\begin{array}{ll}
0 & \mbox{if } \lambda_{n,\ell,j} = \lambda_{n',\ell',j'} \; , \\
1 & \mbox{otherwise} \; ,
\end{array}
\right. 
\end{align}
we can write
\begin{widetext}
\begin{align}
M_{n,\ell,j}^{(1)} (t) =  & \sum_{n'} c^+_{n,n',\ell-1} \psi_{n',\ell-1,j}^* \left[ (1-\chi_{n,\ell,j}^{n',\ell-1,j}) t e^{-\lambda_{n,\ell,j}t}  + \chi_{n,\ell,j}^{n',\ell-1,j} 
\dfrac{e^{-\lambda_{n',\ell-1,j}t}-e^{-\lambda_{n,\ell,j}t}}
{\lambda_{n,\ell,j}-\lambda_{n',\ell-1,j}}   \right] \nonumber \\
& + \sum_{n'} c^-_{n,n',\ell+1} \psi_{n',\ell+1,j}^* \left[ (1-\chi_{n,\ell,j}^{n',\ell+1,j}) t e^{-\lambda_{n,\ell,j}t}  + \chi_{n,\ell,j}^{n',\ell+1,j} 
\dfrac{e^{-\lambda_{n',\ell+1,j}t}-e^{-\lambda_{n,\ell,j}t}}
{\lambda_{n,\ell,j}-\lambda_{n',\ell+1,j}}   \right] \; ,
\end{align}
and
\begin{align}
M_{n,\ell,j}^{(2)} (t) = e^{-\lambda_{n,\ell,j}t} & \sum_{n'} \sum_{n''} \bigg[  
c^+_{n,n',\ell-1}  c^+_{n',n'',\ell-2} \psi_{n'',\ell-2,j}^*  F_{n,\ell,j}^{n',\ell-1,n'',\ell-2}(t) + c^+_{n,n',\ell-1}  c^-_{n',n'',\ell} \psi_{n'',\ell,j}^*  F_{n,\ell,j}^{n',\ell-1,n'',\ell}(t) \nonumber \\
& + c^-_{n,n',\ell+1}  c^+_{n',n'',\ell} \psi_{n'',\ell,j}^*  F_{n,\ell,j}^{n',\ell+1,n'',\ell}(t) + c^-_{n,n',\ell+1}  c^-_{n',n'',\ell+2} \psi_{n'',\ell,j}^*  F_{n,\ell,j}^{n',\ell+1,n'',\ell+2}(t)
\bigg] \, ,
\end{align}
with
\begin{align}
F_{n,\ell,j}^{n',\ell',n'',\ell''}(t) & = (1-\chi_{n',\ell',j}^{n'',\ell'',j}) \Bigg[
(1-\chi_{n,\ell,j}^{n',\ell',j}) \frac{t^2}{2} + \chi_{n,\ell,j}^{n',\ell',j} \dfrac{1 + e^{(\lambda_{n,\ell,j}-\lambda_{n',\ell',j})t} (\lambda_{n,\ell,j}t-\lambda_{n',\ell',j}t-1)  }{(\lambda_{n,\ell,j}-\lambda_{n',\ell',j})^2}
\Bigg] \nonumber \\
& \qquad + \chi_{n',\ell',j}^{n'',\ell'',j}  \Bigg[
(1-\chi_{n,\ell,j}^{n',\ell',j}) \dfrac{e^{(\lambda_{n,\ell,j}-\lambda_{n'',\ell'',j})t} -1 - \lambda_{n,\ell,j}t+\lambda_{n'',\ell'',j}t) }{(\lambda_{n,\ell,j}-\lambda_{n'',\ell'',j})^2} \nonumber \\
& \qquad \qquad +
(1-\chi_{n,\ell,j}^{n'',\ell'',j}) \dfrac{e^{(\lambda_{n,\ell,j}-\lambda_{n',\ell',j})t} -1 - \lambda_{n,\ell,j}t+\lambda_{n',\ell',j}t) }{(\lambda_{n,\ell,j}-\lambda_{n',\ell',j})^2} \nonumber \\
& \qquad \qquad+
\chi_{n,\ell,j}^{n',\ell',j} \chi_{n,\ell,j}^{n'',\ell'',j} \dfrac{1}{\lambda_{n',\ell',j}-\lambda_{n'',\ell'',j}} \Bigg( 
\dfrac{e^{(\lambda_{n,\ell,j}-\lambda_{n'',\ell'',j})t}-1}{\lambda_{n,\ell,j}-\lambda_{n'',\ell'',j}}  - \dfrac{e^{(\lambda_{n,\ell,j}-\lambda_{n',\ell',j})t}-1}{\lambda_{n,\ell,j}-\lambda_{n',\ell',j}} 
\Bigg)
\Bigg] \; .
\end{align}

\end{widetext}

%\bibliographystyle{apsrev4-1-title}
%\bibliography{references}


\end{document}